\documentclass[conference]{IEEEtran}
\IEEEoverridecommandlockouts
\usepackage{float}
\usepackage{comment}
\makeatletter
\let\newfloat\newfloat@ltx
\makeatother
\usepackage[utf8]{inputenc}
\usepackage{xcolor}
\usepackage{graphicx}
\usepackage{amsmath}
\usepackage{amsthm}
\usepackage[normalem]{ulem}
\usepackage[flushleft]{threeparttable}
\usepackage{amssymb}
\usepackage{url}
\usepackage{soul}
\usepackage{subfig}
\newcommand\redout{\bgroup\markoverwith
{\textcolor{red}{\rule[.5ex]{2pt}{0.4pt}}}\ULon}
\usepackage[ruled,linesnumbered,vlined]{algorithm2e}
\usepackage{etoolbox}

\newcounter{algoline}

\AtBeginEnvironment{algorithm}{\setcounter{algoline}{0}}
\usepackage{booktabs}
\usepackage{multirow}

\newboolean{short}
\setboolean{short}{true}
\newcommand{\shortOnly}[1]{\ifthenelse{\boolean{short}}{#1}{}}
\newcommand{\onlyShort}[1]{\ifthenelse{\boolean{short}}{#1}{}}
\newcommand{\longOnly}[1]{\ifthenelse{\boolean{short}}{}{#1}}
\newcommand{\onlyLong}[1]{\ifthenelse{\boolean{short}}{}{#1}}

\let\originalleft\left
\let\originalright\right
\renewcommand{\left}{\mathopen{}\mathclose\bgroup\originalleft}
\renewcommand{\right}{\aftergroup\egroup\originalright}

\def\BibTeX{{\rm B\kern-.05em{\sc i\kern-.025em b}\kern-.08em
    T\kern-.1667em\lower.7ex\hbox{E}\kern-.125emX}}
\begin{document}
\title{Machine Learning Assisted Bad Data Detection for High-throughput Substation Communication \vspace{-3mm}}

\author{\IEEEauthorblockN{Suman Sourav\IEEEauthorrefmark{1}\IEEEauthorrefmark{2}, Partha P. Biswas\IEEEauthorrefmark{1}, Vyshnavi Mohanraj\IEEEauthorrefmark{1}, Binbin Chen\IEEEauthorrefmark{1}\IEEEauthorrefmark{2}, Daisuke Mashima\IEEEauthorrefmark{1}}
\IEEEauthorblockA{\IEEEauthorrefmark{1}Singapore University of Technology and Design, Singapore\\ \IEEEauthorrefmark{2}Advanced Digital Sciences Center, Singapore} 
Email:    [suman\_sourav, binbin\_chen]@sutd.edu.sg; [partha.b, vyshnavi.m, daisuke.m]@adsc-create.edu.sg}

\maketitle

\begin{abstract}
Electrical substations are becoming more prone to cyber-attacks due to increasing digitalization. Prevailing defence measures based on cyber rules are often inadequate to detect attacks that use legitimate-looking measurements\onlyLong{or control messages}. 
In this work, we design and implement a bad data detection solution for electrical substations called \emph{ResiGate}, that effectively combines a physics-based approach and a machine-learning-based approach to provide substantial speed-up in high-throughput substation communication scenarios, while still maintaining high detection accuracy and confidence.
While many existing physics-based schemes are designed for deployment in control centers (due to their high computational requirement), ResiGate is designed as a security appliance that can be deployed on low-cost industrial computers at the edge of the smart grid so that it can detect local substation-level attacks in a timely manner.
A key challenge for this is to continuously run the computationally demanding physics-based analysis to monitor the measurement data frequently transmitted in a typical substation.
To provide high throughput without sacrificing accuracy, ResiGate %
uses machine learning to effectively filter out most of the non-suspicious (normal) data and thereby reducing the overall computational load,  allowing efficient performance even with a high volume of network traffic. 
We implement ResiGate on a low-cost industrial computer and our experiments confirm that ResiGate can detect attacks with zero error while sustaining a high throughput.
\end{abstract}

\section{Introduction}
\label{sec:introduction}
Substations are critical nodal points in a power grid that handle the transmission and distribution of power through different components like transformers, switchgears with busbars and circuit breakers, and intelligent electronic devices (IEDs).
In recent years, researchers have studied various cyber threat scenarios for electrical substations and their mitigation strategies (e.g., \cite{gupta2019prevailing,he2016cyber}). \onlyLong{,tabu2019). The digital components in a smart substation communicate among themselves using standard industrial protocols including Modbus TCP, IEC 61850, etc.}Security technologies such as firewall and intrusion detection system (IDS) are being increasingly deployed to protect the substation networks.
Firewall and IDS typically work on a specific set of cyber rules to allow legitimate traffic\onlyLong{ inside a network}. 
While they can flag/block some unauthorized or malicious packets, they may fail to counter advanced attacks that follow the normal communication channel and use legitimate-looking measurements\onlyLong{or control messages } to cause physical damage.

In this work, we design and implement \emph{ResiGate}, which is a physics-based intrusion detection solution for electrical substations. ResiGate's \onlyLong{physics-based} approach could supplement the cyber-based rules in today's intrusion detection systems. 
Specifically, ResiGate checks for existence of false data in the measurement reported by IEDs in a substation.
While physics-based approaches for detecting false measurement data \onlyLong{and malicious control commands }have been studied in the literature~\cite{schweppe1970power,mashima2017artificial,mashima2018securing}, most existing solutions are  designed for deployment in control centers. Such centralized solutions often do not take into consideration the high-frequency measurement data available at substations (that is not reported at a control center due to the limited bandwidth constraints). Additionally, they rely on high-performance servers to handle the computationally onerous physics-based analysis, and also require fast and reliable communication channel. Though physics-based analysis is accurate, it is also computationally demanding which makes its deployment difficult in scenarios with limited computational resources or requiring processing high throughput data. If only periodically sampled data is sent to the control center for detection, an advanced attacker can avoid the detection by launching only short-duration transient attacks at the base station level.

\onlyShort{{To mitigate this and to effectively perform bad data detection at the local substation level, in this paper, we propose to augment the physics-based false data detection with machine learning, whereby the goal of the machine learning is to reduce the amount of computation, by filtering out non-suspicious normal data.}} Specifically, for this filtering ResiGate first scans all measurements using a lightweight ML algorithm, namely gradient boosted decision tree (GBDT). We should also note that Resigate framework is agnostic to ML algorithm, and any advanced scheme can be integrated. The GBDT model for a given substation is trained in advance using diverse sets of normal data and pre-crafted abnormal data. Once trained, the model requires only about 20 milliseconds for online classification of a snapshot of over 100 measurement points in our experimental setup. A small set of data flagged as suspicious will be further validated by physics-based analysis. 
Now, as only a small set of suspicious data needs to undergo physics-based analysis, the overall computational resources required to validate the data would be significantly reduced, providing accelerated bad data detection. Although both ML-based and physics-based IDS solutions for power systems have been proposed in literature (e.g.,~\cite{9042544,singh2020cyber}), to the best of our knowledge, ResiGate is the first fully-functioning research prototype that combines the efficiency of an ML-based approach with the accuracy of physics-based analysis for low-cost deployment in electrical substations. The main contributions of our work are as follows:

\begin{itemize}
    \item We propose a novel design in ResiGate that augments the accurate but computationally demanding physics-based analysis with an efficient machine learning based filtering module. Our design enables ResiGate to operate at the local substation level (as opposed to centrally located control-centers) over a low-cost industrial computer while handling high-throughput of up to 61 snapshots per second. 
    Resigate maintains the confidence and accuracy of a traditional physics-based method while still having low-enough latency and computational demand to be effectively deployed at the sub-station level.
    \item We give an entirely open-source and ready-to-use tool--`ResiGate' whereby we integrate an open source power system simulator (Pandapower~\cite{thurner2018pandapower}) with an open source network monitoring tool (Zeek~\cite{Zeek}). 
    \item We conduct extensive evaluations of ResiGate prototype in various settings of a test network of 3 nearby substations. Our experimental results confirm that ResiGate can accurately detect attacks on measurement data while sustaining high throughput, where some substation-level event triggers multiple IEDs into fast transmission mode.  
\end{itemize}

\section{Literature Review}

Traditionally, IDS solutions use signature-based, behaviour-based or specification-based approaches to detect attacks. They have been used in various cyber-physical systems (CPS), including smart grids~\cite{sun2018cyber,berthier2011specification}. Physics-based methods for securing power grid have also been widely studied. Bad data detection (BDD) algorithms can detect erroneous sensor data and malicious data, as long as the number of affected measurement points is below a certain threshold~\cite{liu2011false, kim2011strategic}. 
A false-data injection attack (FDIA) is when an attacker manipulates measurement data with malicious intentions. One popular work on this front is the study on FDIA using power system state estimation. Power system engineers and operators estimate voltage and phase angle in the power network based on sensor measurements at various locations. Most of the existing works on FDIA focus on detecting attacks for a wide area power grid. On the contrary, in this work, we focus on the development of a physics-based intrusion detection for localized attacks, which happen within a single substation (or a cluster of neighboring substations).

Quite a few recent works made use of ML techniques to develop intelligent IDSes (e.g., \cite{napiah2018compression},  \cite{ren2018edmand}\onlyLong{, and \cite{garcia2016comparative}}). \onlyLong{Goh \textit{et al.}~\cite{7911887} studied an unsupervised approach for detecting cyber attacks on cyber-physical systems. }In the context of power systems, Hink \textit{et al.}~\cite{6900095} demonstrated the use of several different ML algorithms for the classification of normal events, cyber-attacks and natural disturbances. A decision-tree based algorithm was suggested by Adhikari \textit{et al.}~\cite{7805317} for the development of binary-class and multi-class classification models. %
More recently, the authors in~\cite{9042544} proposed hybrid intrusion detectors for detecting unknown and stealthy cyber-attacks. %
In general, an ML-based approach is more efficient than physics-based approach for online detection. However, it is difficult to provide assurance about the accuracy of the ML-based approach. When applying an ML-only approach to our setting, where a large number (e.g., several hundreds) of substations need to be continuously monitored (for years or decades), even a small error rate \onlyLong{(e.g., $<0.1\%$)} can accumulate to a large number of false positive and false negative data. 

The application of ML algorithms to detect FDIA is widespread. As evident from the recent survey articles (\cite{musleh2019survey, cui2020detecting}), methods such as support vector machine (SVM), extreme learning machine (ELM), decision tree (DT), deep belief network (DBN), etc., have been extensively used to detect FDIA. Our approach is different from the previous works in the sense that we use the ML algorithm as the facilitator that accelerates and aids the physics-based IDS in detecting bad data among the stream of data packets.

\section{Threat Model}
\label{threatmodel}

Substations are critical entities in the power grid. Intrusion into a substation not only can affect the region directly connected to that substation, but also may lead to cascading events in certain scenarios leading to cause catastrophic failure of the grid\onlyShort{.}
A substation usually has a local control station where operator and engineering workstations are located.  
In addition, it is monitored and sometimes controlled remotely from a control center. An attacker can enter the local control station via the gateway of the substation, as happened in the Ukraine attack case~\cite{case2016analysis}. Another point of entry could be the engineering computer which is connected to IEDs to modify any protection settings. A malware can be installed in the IEDs, similar to what has been done with the PLCs in the Stuxnet attack~\cite{farwell2011stuxnet}.
An attacker may also implant malware in IEDs through their supply chains. Having gained a foothold inside the substation, the attacker can use the compromised devices to inject malicious measurement data\onlyLong{ or commands}. Malicious information could also come from the compromised control center (or man-in-the-middle attack).

In our case studies, we assume that the attacker is not powerful enough to launch stealthy FDIAs in the substation. To launch a stealthy FDIA the attackers need to control sufficient measurement points so that the manipulated states still satisfy the consistency checking of non-advanced state estimators. Usually, in order to mount such an attack, an attacker needs the complete information of the power grid topology and configuration. In practice, this is often not the case. Here, we focus on faster detection for countering ``random" (non-intelligent) FDIA that can target multiple measurements simultaneously. We also assume that the attacker cannot compromise the deployed IDS or its visibility to all the traffic inside the substation network. In this context, do note that other advanced attack detection modules can be integrated into the ResiGate framework. We plan to investigate the upgrading of the physics-analysis component to an advanced state estimator that can handle stealthy FDIA in a future work.

\section{ResiGate Design}
\label{Design}

In this section, we will first lay our design objectives. Thereafter, we present an overview of ResiGate's design to achieve the objectives. 
\subsection{Design Objectives}
To address the various shortcomings of existing IDS solutions for power systems, %
we aim to design an edge-based ResiGate IDS to achieve desired performance in the following four key aspects:
 
\begin{itemize}
\item Security: ResiGate shall be capable of detecting local attacks launched in individual substations as described in Section~\ref{threatmodel}. 
\item Accuracy: The accuracy of detection of the IDS shall be very high so that no attack can evade the detection method. %
Online state estimation of power system can detect FDIA with very high accuracy. We utilize such highly reliable characteristics of physics-based analysis to attain near-zero error in detection.
\item Performance: Here we are mainly concerned with supporting high throughput processing of measurement traffic for a medium to large size substation (or a cluster of neighboring substations), which can contain dozens of IEDs with hundreds of measurement points. ResiGate should be able to process the data even under the most demanding situation. We use ML-based filtering to improve performance, without much compromise of accuracy.

\item Deployment: To allow large-scale deployment of our solution at the edge of the power grid (i.e., in substations), ResiGate should be deployable on low-cost commercial off-the-shelf (COTS) industrial computers. 
\end{itemize}

\subsection{ResiGate Overall Design}

\begin{figure}
 \centering
    \includegraphics[width=0.85\linewidth]{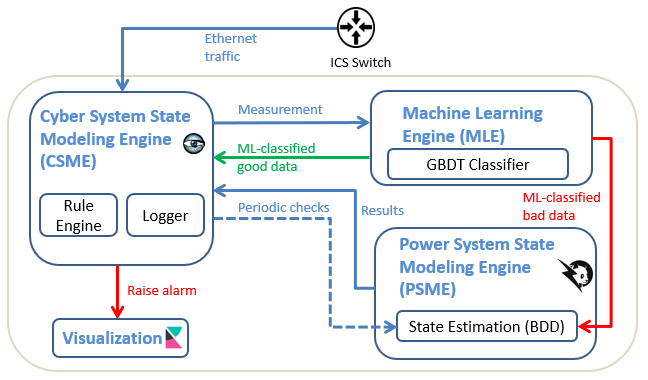}
  \caption{ResiGate architecture\vspace{-3mm}}
  \label{fig:Resigate}
\end{figure}

The ResiGate architecture, shown in Fig.~\ref{fig:Resigate}, consists of three major components; Cyber System State Modeling Engine (CSME), Power System State Modeling Engine (PSME) and  Machine Learning Engine (MLE).

The \textit{Cyber System State Modeling Engine} (CSME) is required to have functionalities of monitoring the network traffic. In ResiGate, we adopt the open source tool Zeek~\cite{Zeek} for network monitoring. Zeek offers protocol parsing capabilities as well as the options to create and customize security policies. One can set up IDS rules using the Zeek framework. Suitable parsers inside CSME parse the packets that ResiGate obtains via the mirror port of the industrial switches inside substations.

The \textit{Power System State Modeling Engine} (PSME) is built upon an open-source power system simulator called Pandapower~\cite{thurner2018pandapower}. 
Pandapower can perform operations such as power flow analysis and state estimation on the network model built using its internal or user-defined power system component library. It is worthwhile to mention that Pandapower is one proof-of-concept based on readily available open source solutions. The framework can be used with other FDIA detection schemes. We also note that the development of a mechanism to counter advanced, sophisticated FDIA attacks is orthogonal to the scope of this paper.

The \textit{Machine Learning Engine} (MLE) implements an ML algorithm and processes the input data using a pre-trained substation-specific model. We select the gradient boosted decision
tree (GBDT) algorithm, a well-known supervised machine learning model that has been widely used for classification and regression (e.g.,~\cite{hu2019advanced,cheng2020powernet}) purposes.

The PSME in our IDS is highly accurate in detecting malicious data injection attacks and malicious commands as it uses substation-specific physical model and up-to-date measurement and status information. Using PSME as the final decision maker achieves the security and accuracy objectives. However, ResiGate does not send all network packets to PSME as PSME has high computational overhead. To enhance the throughput performance, network packets are filtered by MLE and only suspicious packets are sent to PSME. The ML algorithm (in MLE) also has a reasonably high accuracy as it is trained on substation-specific model and load flow data. During normal operation (i.e., in the absence of an attack), the PSME does not participate in the IDS operation except for being periodically invoked in case some bad data evade the ML classifier (i.e., false negative). CSME coordinates the entire cycle of operation after receipt of network packets. Thus, the overall process helps ResiGate achieve high throughput while maintaining high accuracy. The combination of PSME and MLE also allows ResiGate to work on a low-cost industrial computer that can be deployed as an edge security appliance.

\section{ResiGate Implementation}

We have implemented ResiGate on an industrial computer from DA Vision (Model: MPCX-H110)  %
with 16 GB DDR4 SO-DIMM RAM and 4 cores of 2.8 GHz processors (Intel Core i7-6700T with 8MB of cache), running on Ubuntu 18.04 operating system. Our implementation is based on Zeek version 2.6 and Pandapower version 2.1.0. \onlyLong{Demo videos of ResiGate are available on our project website~\cite{ResiGate}, where the setup and key features of ResiGate are explained. In the following, we briefly present our implementation to process the measurement data and commands in ResiGate.}

\subsection{The Processing of Measurement Data}

In ResiGate, CSME uses relevant protocol analyzers to parse the packet contents. 
Upon parsing of a packet that contains measurement data, a Zeek event containing the key and value of the data is generated. 
In the event handler, ResiGate maintains a table of the latest power and voltage readings from all IEDs. A snapshot of this table is periodically sent to the MLE for analysis.

The MLE normalizes each snapshot of datapoints received from CSME to a suitable format for evaluation. If a \onlyLong{particular} snapshot \onlyLong{of the data} is classified as suspicious by MLE, it will be sent to the PSME. PSME internally maintains the static configuration (including the topology and specifications of various elements such as buses, switches and transformers) of the power system.

\subsection{Machine Learning Meets Physics-based Analysis}

In power system state estimation, the voltage and voltage angles of various system buses are estimated based on available measurements obtained from several field metering instruments. 
The state estimation accounts for the measurement errors by performing a regression analysis with all the available measurement data (including a certain level of redundancy).
Mathematically, the relationship between state variables and measurements is given by~\cite{schweppe1970power,liu2011false}: 
\begin{equation}\label{SE}
    z = h(x)+e
\end{equation}
where $z$ is the measurement vector of $m$ measurements, $x$ is the state vector of $n$ states, $e$ is the measurement error vector, and $h(.)$ is a set of non-linear functions of state vector. In state estimation, the state vector $x$ is estimated using weighted least square (WLS) method by solving the following equation: 
\begin{equation}\label{SEst}
    \hat{x} = \text{arg} \min_{x}[z-h(x)]^TW^{-1}[z-h(x)]
\end{equation}
where $W$ is the matrix of reciprocals of the measurement error variances.
The solution to Equation (\ref{SEst}) is determined through an iterative approach. To find bad data in measurements, traditionally, chi-squared test is carried out on the performance index defined as:
\begin{equation}\label{PERFIDX}
    J(\hat{x}) = \sum_{i=1}^m\frac{[z_i-h_i(\hat{x})]^2}{\sigma_i^2}
\end{equation}
where $\sigma_i^2$ is the variance of $i$-th measurement $(1\leq i\leq m)$. With the assumptions that state variables are mutually independent and the measurement errors follow normal distribution, it can be proved that $J(\hat{x})$ follows $\chi_{(m-n)}^2$ distribution, where $(m-n)$ is the degree of freedom. The threshold of identifying bad data is $\chi_{(m-n),p}^2$, which is the value of the chi-squared distribution for the probability $p$ and $(m-n)$ degree of freedom.  If $J(\hat{x})$ exceeds this threshold, bad data is suspected in measurements.

The inputs to the Pandapower state estimator are a number of measurements in the specified format. To identify and eliminate bad data among measurements, a wrapper function \textit{chi2\_analysis}, i.e., `Chi-squared' analysis is developed in Pandapower. The function works on the classical principle of state estimation as discussed above. As the process involves iterative approach of estimating system state, the state estimation bad data detector has inherently high latency. Therefore, it becomes necessary to accelerate the process of bad data detection in systems where messages are exchanged at fast rates.

Before sending the measurement data to Pandapower bad data detector (BDD), we use ML algorithm GBDT to filter the stream of measurement values. We select GBDT because it works well on our classification datasets and serves the purpose of boosting the physics-based IDS. We leave the comparative performance study among few other ML algorithms for the future. Bad data may occasionally pass the ML algorithm filtering stage. To counter such cases of wrong classification of bad data by the ML algorithm, PSME is invoked periodically even if not alerted by the MLE.

Another way to deal with these false negatives would be to use ML classifiers that give a score (rather than a binary result). Depending on the requirements, a threshold can be set to pass suspicious data points to PSME. ML models can also be trained in a way where the loss function penalizes false negatives much more than the false positives. Lowering the threshold, would reduce false negatives, but would also lower performance (in terms of computational-load/latency). Also, ensemble models where the threshold is determined in terms of the number of ML models that raise alarm, can be used.

\section{Case Studies and Evaluations}
\label{evaluations}
We evaluate ResiGate in detail for false data injection attacks under various configurations of a test system. \onlyLong{We also present a case of detecting malicious commands in the substation by ResiGate.} 

\begin{figure}[t!]
  \centering
    \includegraphics[width=0.75\linewidth]{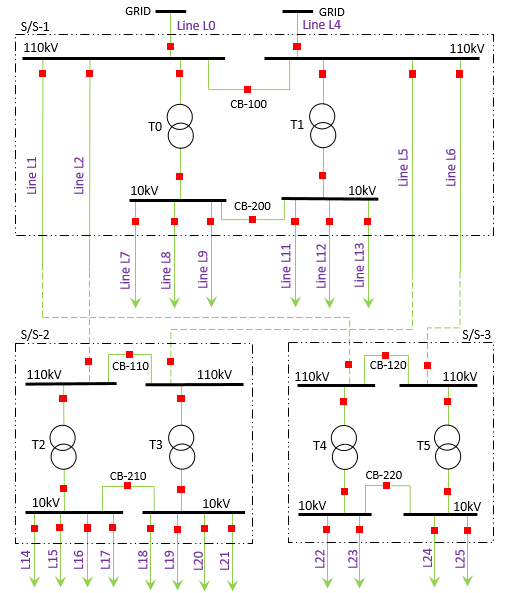}
  \caption{One-line diagram of the test network\vspace{-3mm}}
  \label{fig:3sub}
\end{figure}

\subsection{Test System}

In our study, we consider a 3-substation model as shown in Fig.~\ref{fig:3sub}. As seen in the diagram, substation S/S-1 receives feed from the main grid. Substations S/S-2 and S/S-3 are connected to substation S/S-1 with redundant feeders (lines). In each of these substations, two transformers step down the voltage to the required level. Under a normal configuration, all loads and equipment are in operation, all switches except the bus section switches in substations S/S-2 and S/S-3 (i.e., CB-110, CB-210, CB-120 and CB-220) are closed. There are 32 IEDs in these substations, with at least one IED in each feeder (line) for control and monitoring purposes. The interconnection of IEDs and a more detailed one-line diagram of single substation can be found in~\cite{biswas2019synthesized}. The loads are connected at the lines terminated with the arrows. Therefore, a total of 18 load points (active/reactive power consumers) are present in the 3 substations. The substations are located in close geographical location from one another and the set up represents a localized network. We use a single ResiGate device to process traffic from all the substations and evaluate its performance.

\subsection{Generating Datasets for the Machine Learning Algorithm}

We create synthesized data to train and test the ML model. A power system can have different configurations, e.g., switching off of some loads, outage of a line, isolation of an equipment for maintenance, etc. Of course, by training on all the possible configurations results in higher accuracy as shown in \cite{5128907,5504793}, however, oftentimes it is not practical to train the ML model for all possible configurations. In our experiments, we train the ML model for only normal configuration of the test system containing both normal and corrupt measurements (see Fig.~\ref{fig:3sub}), where we vary the loads and use the results obtained from power flow study. To show the robustness of this approach, we test the ML model with different configurations for which it is not trained with. As shown later, despite the slightly unbalanced training data, the ML model is robust delivers good accuracy for the different configurations. %

To implement the ML model GBDT, we use `scikit-learn' python library for tuning the relevant hyper-parameters using grid based search. Some of the key parameters include the number of boosting stages `\emph{n\_estimators}', the learning rate `\emph{learning\_rate}' that determines the contribution from each tree, and the maximum depth of the individual regression estimators `\emph{max\_depth}' that sets a limit on the number of nodes in the tree. %
The search range in the first step are $n\_estimators$: \{100, 200, 250, 300, 350, 400, 500\}, $max\_depth$: \{1, 2, 3, 4, 5\} and $learning\_rate$: \{0.001, 0.01, 0.1, 1\}. \onlyLong{The }$n\_estimators$\onlyLong{ parameter} is tuned further within the search range of [200,300] in steps of 10 for our case studies and evaluations.

\begin{table*}
\centering
\caption{Results of ML and Pandapower BDD (with up to $\pm{0.5\%}$ noise)}
\label{tbl:MLoutputmulti}
\begin{threeparttable}
\begin{tabular}{|p{12mm}|p{35mm}|p{12mm}|p{12mm}|p{12mm}|p{12mm}|p{12mm}|p{12mm}|}\hline
\textbf{Settings} & \textbf{Description} & \textbf{ML True negative} & \textbf{ML True positive} & \textbf{ML False negative} & \textbf{ML False positive} & \textbf{ML \newline Precision} & \textbf{BDD \newline Precision}\\ \hline
Normal & All in operation & 2000 & 500 & 0 & 0 & 1 & 1 \\ \hline
Config. 1 & Line L1 out, CB-120 closed & 1988 & 500 & 0 & 12 & 0.9766 & 1 \\ \hline
Config. 2 & Line L2 out, CB-110 closed & 1975 & 500 & 0 & 25 & 0.9524 & 1 \\ \hline
Config. 3 & Line L8 off & 1996 & 500 & 0 & 4 & 0.9921 & 1 \\ \hline
Config. 4 & Line L11 off & 1997 & 500 & 0 & 3 & 0.9940 & 1 \\ \hline
Config. 5 & Line L14 off & 1985 & 500 & 0 & 15 & 0.9709 & 1 \\ \hline
Config. 6 & Line L25 off & 1983 & 500 & 0 & 17 & 0.9671 & 1 \\ \hline
\end{tabular}
\begin{tablenotes}\footnotesize
\item Note: The recall of ML and BDD remains at 1 for all settings
\end{tablenotes}
\end{threeparttable}
\end{table*}

\subsubsection{Training Datasets}
The loads, i.e., active ($p$) and reactive power ($q$) consumers, connected to a substation are variable under normal circumstances. \onlyLong{Ref.~\cite{esmalifalak2014detecting} considered the load variation within the range from 90\% to 110\% of the nominal rating. We avoid the excessive loading scenario and }We consider the normal range from 70\% to 100\%. In case of an attack, the attacker may falsify the $p$ and $q$ readings of the loadpoints in addition to the voltage ($V$) at various system buses. The attacker would try to manipulate the readings in a way that the change would impact the physical system but it would not trigger the simple threshold-based anomaly detectors. To simulate such behavior, we choose a multiplication factor within the range from 0.7 to 1.3 to corrupt the original power flow readings. In summary, we adopt the following steps to generate the training datasets for the ML algorithm:

\begin{itemize}
    \item We vary the loads randomly in the network from 70\% to 100\% of their nominal rating (both $p$ and $q$ values) to create the desired number of load datapoints.
    \item We run power flow using Pandapower power simulator for all the load datapoints.
    \item From power flow results of each load datapoint, we record $V$, $p$ and $q$ for all buses except for the buses where loads are directly connected; we also record $p$ and $q$ for all lines and transformers.
    \item To create corrupt dataset, we corrupt 20\% of the total load datapoints and in each corrupt datapoint, we corrupt 30\% of the readings by multiplying with a randomly generated corruption factor ranging from 0.7 to 1.3. For example, if we generate 1000 load datapoints with each having 40 measurements (readings), we create 200 corrupt datapoints and in each of these corrupt datapoints, we alter 12 measurements by multiplying with 12 randomly generated corruption factors within the range $[0.7,1.3]$.  
    \item We label the legitimate and corrupt data for the purpose of classification.
\end{itemize}

\subsubsection{Test Datasets}
Test datasets for the ML model in IDS are created in a way similar to the training datasets. The point to note here is that we create test datasets for many different configurations of the multi-substation experimental setup. We also introduce a noise of upto $\pm{0.5\%}$ in the test datasets to account for measurement errors.

We also introduce different noise levels in selected case studies of test datasets to account for the measurement errors in state estimation. The maximum noise level considered in our test datasets is $\pm{1.5\%}$. As per the standard IEEE C57.13.6, the high accuracy classes of instrument transformers such as $0.15$S, $0.3$S, $0.5$S, etc. have maximum errors of $\pm{0.15\%}$, $\pm{0.3\%}$ and $\pm{0.5\%}$, respectively. Further, the standard ANSI C12.20 establishes the metering instrument accuracy classes of 0.2, 0.5, etc. Another commonly deployed metering instrument in the electrical network is the phasor measurement unit (PMU) whose accuracy is defined by the total vector error (TVE)~\cite{lixia2010accuracy} and the same shall be limited to 1\%. Therefore, the maximum noise level of up to $\pm{1.5\%}$ can be considered as a reasonable assumption for the case studies. 

\addtolength{\topmargin}{0.05in}
\subsection{Evaluation Results}

\subsubsection{Results for different system configurations}
As mentioned before, there are 32 IEDs in these substations and each of them sends a continuous stream of packets. Each network packet carries line or bus measurements from an IED, containing active power, reactive power and voltage data. In our study cases, we get a total of 124 measurement points from all IEDs. The measurements are being transmitted by the IEDs periodically. Depending on whether there are events, one IED will send out one update or 10 updates per second. ResiGate can take multiple snapshots per second to support timely detection. All the packets are checked and subsequently validated as necessary by ResiGate.
We train the ML model with the data from power flow study. The training data used in this case contains 10000 different measurement snapshots (datapoints). We create test datasets of size 2500 each for different configurations of the network. These test datasets are adulterated with noise up to $\pm{0.5\%}$ to account for the measurement errors. The various configurations and the output of our IDS are summarized in Table~\ref{tbl:MLoutputmulti}. To elaborate, `Config. 1' in the table refers to the case when the connecting line $L1$ is out of service and the left section of substation S/S-3 is fed by closing the breaker CB-120. Using this configuration, we run power flow to generate the required test datapoints by varying the connected loads, and adding up to $\pm{0.5\%}$ noise to the power flow results. `Config. 2' can be interpreted in a similar manner. Line outages in `Config. 3' to `Config. 6' refer to the non-supply of power to the loads connected to the respective lines. In each case, 20\% of the datapoints (i.e., 500 nos.) are corrupted with the randomly generated factors within the range $[0.7,1.3]$.

With reference to the ML results in Table~~\ref{tbl:MLoutputmulti}, `true positive' ($TP$) signifies correct classification of bad data while `true negative' ($TN$) denotes right classification of the good data. On the contrary, `false negative' ($FN$) indicates wrong classification of bad data and `false positive' ($FP$) is for incorrect classification of good data as bad data. Precision is the ratio defined by $TP/(TP+FP)$, while recall is defined by $TP/(TP+FN)$. As observed from the tabulated results, the GBDT algorithm catches all corrupt data correctly for all configurations of the system. In some cases, good data is falsely classified as bad data. Further validation of the corrupt data flagged by the ML method is performed using Pandapower BDD. The Pandapower BDD checks all $TP$ and $FP$ data identified by the ML algorithm. The $TP$ data fail to pass the Pandapower state estimator based BDD, thus, reinstating their classification as bad data. The $FP$ data pass the detection test to revert their original state of legitimate data. We get very few $FN$, totaling to only 3 over all the cases with different configurations. As ML algorithm classifies these bad data as good data, the Pandapower BDD would not act on these. To counter such cases, we run the Pandapower BDD periodically over randomly selected snapshot samples even if no flag is raised by the ML algorithm. 
Such periodic checks could serve as a second line of defence to detect the small percentage of bad data where the ML algorithm fails to identify.
To demonstrate the accuracy of Pandapower BDD, we also test all datapoints in all the cases (for different system configurations) with Pandapower BDD. The Pandapower BDD precision and recall are found to be 1 (with zero false positive and zero false negative), as reported in Table~\ref{tbl:MLoutputmulti}. However, the BDD alone takes much longer time to execute and would not be able to support real-time high throughput scenarios\onlyLong{ triggered by some substation level event}. %
The filtering of data packets by the ML reduces the burden from BDD and accelerates the entire process. Therefore, our approach to augment the physics-based IDS by incorporating ML works effectively and more efficiently in the end-to-end implementation, helping us to fulfil our design objectives discussed in Section~\ref{Design}.

\begin{figure}
  \centering
    \includegraphics[width=0.7\linewidth]{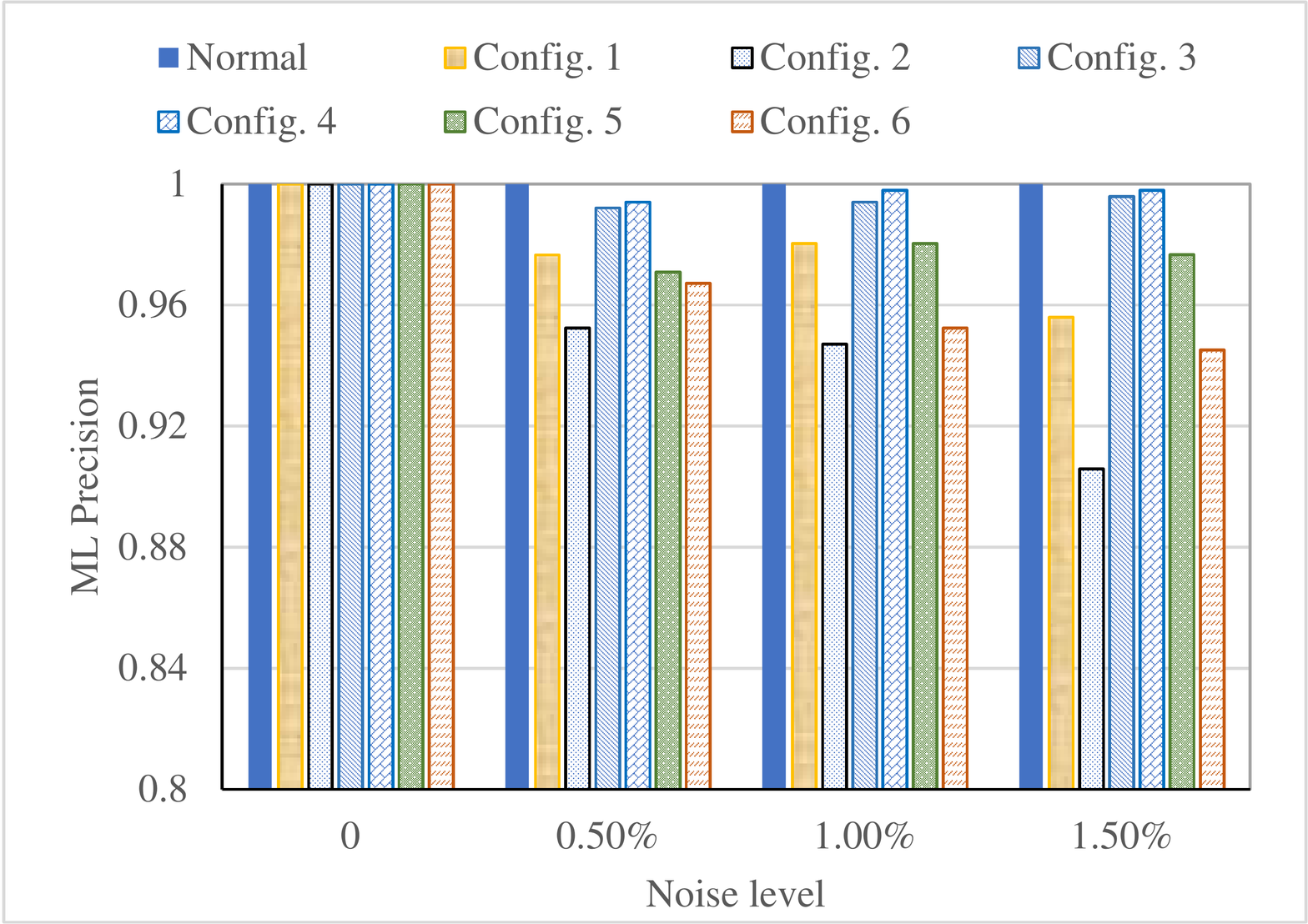}
    \caption{\small Precision of the ML algorithm for different configurations and noise levels. 
    (Recall is almost 1 in the settings, plot not shown)
    \vspace{-3mm}}
    \label{fig:MLprecision}
\end{figure}

\subsubsection{Results for different noise levels}
In the previous section, we discussed our IDS results for measurements with noise level of up to $\pm{0.5\%}$. We test our IDS for different noise levels in measurement datasets. Fig.~\ref{fig:MLprecision} shows the precision of the ML algorithm for all the system configurations at different noise levels. As expected, the precision drops when the noise is high; nonetheless, it is still 0.9 or higher. We run the Pandapower BDD independently in all cases (with different noise levels), and it accurately detects all bad data with no false positive or false negative. This reiterates the usefulness of the highly accurate physics-based IDS whose performance is enhanced with the application of an ML algorithm.

\subsubsection{System latency and throughput}
\label{latency}
\begin{figure}[t!]
  \centering
    \includegraphics[width=0.75\linewidth]{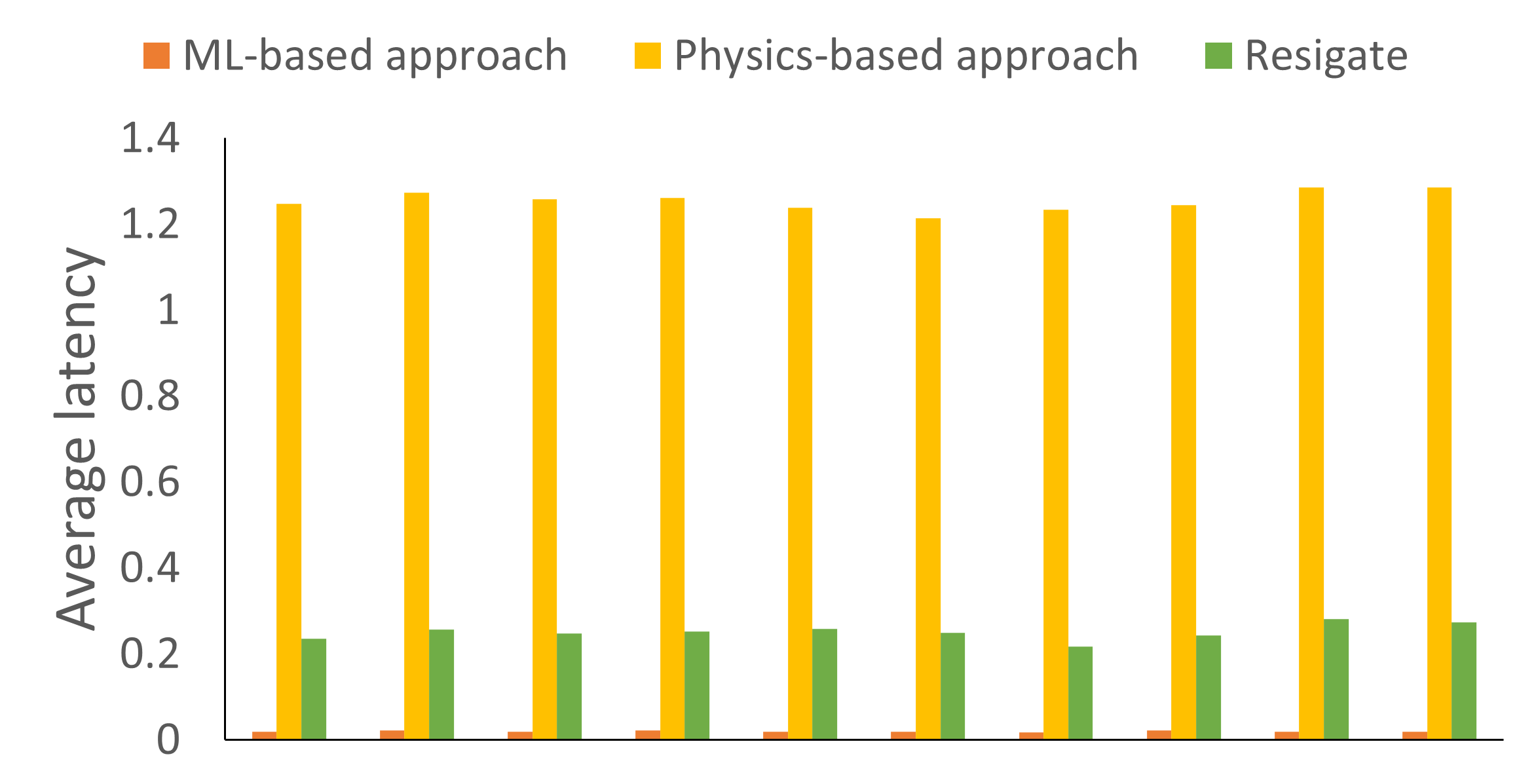}
  \caption{\small Plot showing the average latency per snapshot (in seconds) of the ML-based approach and the physics-based approach when run in a stand-alone fashion while comparing them to Resigate for 10 different instances  where 20\% of snapshots are corrupted.\vspace{-3mm}}
  \label{fig:latency}
\end{figure}

We define the latency (i.e., execution time) of processing a snapshot of measurement data by ResiGate as the time duration between the moment the CSME sends out the snapshot (to MLE) and the moment the CSME is informed about the decision (with or without invoking PSME). The latency can differ significantly depending on whether the PSME (i.e., Pandapower BDD) is involved in the process. 
When there is no attack (i.e., PSME is not invoked), based on our measurements, the typical execution latency of ResiGate is less than 0.02 second per snapshot. The latency increases to around 1.25 seconds per snapshot when the PSME is called on. We also evaluate the case where 20\% of snapshots contain corrupt data (PSME validation is needed). ResiGate takes about 0.25 second of processing time per snapshot on average, which is roughly 1.25 seconds $\times$ 20\%. When running the MLE and the PSME in a stand-alone fashion as the ML-based and the physics-based approach respectively, we see from Figure \ref{fig:latency} that the latency required in those cases is also similar. Even though Resigate's average latency is slightly higher than the pure ML-based approach, the combination of MLE and PSME in Resigate increases the confidence and accuracy of the method while still having low-enough latency to be effectively deployed at a sub-station.

The average processing latency per snapshot directly affects the ResiGate's throughput. As a concrete throughput performance target, we refer to some typical traffic characteristics for an IED transmitting IEC61850 GOOSE messages. Typically, an IED transmits at a frequency of 1 packet per second during normal operation~\cite{en12122272}. %
When there is a status change or upon detection of some important substation events, a faster transmission rate is activated by the affected IED for a few cycles. The interval between packets starts with a small value, and is then exponentially increased until the transmission rate for normal operation is reached~\cite{Hoyos2012ExploitingTG}. In a typical setup of such event-triggered fast transmission mode, the initial fast re-transmission interval can be as low as 1 millisecond (ms), and the interval of subsequent packets would be doubled (i.e., from 1 ms to 2 ms, 4 ms, \dots , up to around 1 second). This results in a higher throughput requirement of around 10 packets per second in the first second upon a substation event. ResiGate should be able to support the high throughput requirement in the most demanding case where all IEDs enter into the fast transmission mode simultaneously. Our evaluation shows that the ResiGate prototype can process up to 61 snapshots per second for attack-free data in the test network we studied. If ResiGate encounters continued attack packets in large percentage, its throughput will drop as PSME will be invoked more often. However, when the first few batches of attack data are detected, the substation should already be put into an emergency mode\onlyLong{, and ResiGate's main function should change to recording (instead of classifying) the data for post-attack analysis}. Hence, the limited throughput supported by PSME is not critical under such a scenario.

\section{Conclusion}
In this work, we design and implement ResiGate, a physics-based edge IDS solution for power system, the performance of which is enhanced with machine learning. To implement the IDS, we combine an open source network monitoring tool (Zeek) with an open source power simulator (Pandapower). 
The performance of our proposed IDS solution highly depends on the accuracy of the ML algorithm. Therefore, further studies using different datasets and ML algorithms would be performed in the future. The IDS solution also needs to be tested for large transmission networks.

\bibliographystyle{IEEEtran}
\begin{small}
\bibliography{ref1}

\begin{thebibliography}{10}
\providecommand{\url}[1]{#1}
\csname url@samestyle\endcsname
\providecommand{\newblock}{\relax}
\providecommand{\bibinfo}[2]{#2}
\providecommand{\BIBentrySTDinterwordspacing}{\spaceskip=0pt\relax}
\providecommand{\BIBentryALTinterwordstretchfactor}{4}
\providecommand{\BIBentryALTinterwordspacing}{\spaceskip=\fontdimen2\font plus
\BIBentryALTinterwordstretchfactor\fontdimen3\font minus
  \fontdimen4\font\relax}
\providecommand{\BIBforeignlanguage}[2]{{%
\expandafter\ifx\csname l@#1\endcsname\relax
\typeout{** WARNING: IEEEtran.bst: No hyphenation pattern has been}%
\typeout{** loaded for the language `#1'. Using the pattern for}%
\typeout{** the default language instead.}%
\else
\language=\csname l@#1\endcsname
\fi
#2}}
\providecommand{\BIBdecl}{\relax}
\BIBdecl

\bibitem{gupta2019prevailing}
A.~Gupta, A.~Anpalagan, G.~H. Carvalho, L.~Guan, and I.~Woungang, ``Prevailing
  and emerging cyber threats and security practices in iot-enabled smart grids:
  A survey,'' \emph{JNCA}, vol. 132, pp. 118--148, 2019.

\bibitem{he2016cyber}
H.~He and J.~Yan, ``Cyber-physical attacks and defences in the smart grid: a
  survey,'' \emph{IET Cyber-Physical Systems: Theory \& Applications}, vol.~1,
  no.~1, pp. 13--27, 2016.

\bibitem{schweppe1970power}
F.~C. Schweppe and J.~Wildes, ``Power system static-state estimation, {P}art
  {I}: Exact model,'' \emph{IEEE TPAS}, no.~1, pp. 120--125, 1970.

\bibitem{mashima2017artificial}
D.~Mashima, P.~Gunathilaka, and B.~Chen, ``Artificial command delaying for
  secure substation remote control: Design and implementation,'' \emph{IEEE
  Transactions on Smart Grid}, vol.~10, no.~1, pp. 471--482, 2017.

\bibitem{mashima2018securing}
D.~Mashima, B.~Chen, T.~Zhou, R.~Rajendran, and B.~Sikdar, ``Securing
  substations through command authentication using on-the-fly simulation of
  power system dynamics,'' in \emph{IEEE SmartGridComm}, 2018, pp. 1--7.

\bibitem{9042544}
V.~K. {Singh}, E.~{Vaughan}, J.~{Rivera}, and A.~{Hasandka}, ``{HIDES}: Hybrid
  intrusion detector for energy systems,'' in \emph{IEEE TPEC}, 2020, pp. 1--6.

\bibitem{singh2020cyber}
V.~K. Singh and M.~Govindarasu, ``Cyber kill chain-based hybrid intrusion
  detection system for smart grid,'' in \emph{Wide Area Power Systems
  Stability, Protection, and Security}.\hskip 1em plus 0.5em minus 0.4em\relax
  Springer, 2020, pp. 571--599.

\bibitem{thurner2018pandapower}
L.~Thurner, A.~Scheidler, F.~Sch{\"a}fer, J.~Menke, J.~Dollichon, F.~Meier,
  S.~Meinecke, and M.~Braun, ``Pandapower-an open-source python tool for
  convenient modeling, analysis, and optimization of electric power systems,''
  \emph{IEEE TPWRS}, vol.~33, no.~6, pp. 6510--6521, 2018.

\bibitem{Zeek}
\BIBentryALTinterwordspacing
``Zeek-ids,'' May 2020. [Online]. Available: \url{https://zeek.org/}
\BIBentrySTDinterwordspacing

\bibitem{sun2018cyber}
C.-C. Sun, A.~Hahn, and C.-C. Liu, ``Cyber security of a power grid:
  State-of-the-art,'' \emph{JEPE}, vol.~99, pp. 45--56, 2018.

\bibitem{berthier2011specification}
R.~Berthier and W.~Sanders, ``Specification-based intrusion detection for
  advanced metering infrastructures,'' in \emph{IEEE PRDC}, 2011, pp. 184--193.

\bibitem{liu2011false}
Y.~Liu, P.~Ning, and M.~Reiter, ``False data injection attacks against state
  estimation in electric power grids,'' \emph{ACM Transactions on Information
  and System Security (TISSEC)}, vol.~14, no.~1, pp. 1--33, 2011.

\bibitem{kim2011strategic}
T.~T. Kim and H.~V. Poor, ``Strategic protection against data injection attacks
  on power grids,'' \emph{IEEE TSG}, vol.~2, no.~2, pp. 326--333, 2011.

\bibitem{napiah2018compression}
M.~N. Napiah, M.~Y. I.~B. Idris, R.~Ramli, and I.~Ahmedy, ``Compression header
  analyzer intrusion detection system ({CHA-IDS}) for 6{L}o{WPAN} communication
  protocol,'' \emph{IEEE Access}, vol.~6, pp. 16\,623--16\,638, 2018.

\bibitem{ren2018edmand}
W.~Ren, T.~Yardley, and K.~Nahrstedt, ``Edmand: Edge-based multi-level anomaly
  detection for {SCADA} networks,'' in \emph{IEEE SmartGridComm}.\hskip 1em
  plus 0.5em minus 0.4em\relax IEEE, 2018, pp. 1--7.

\bibitem{6900095}
R.~C. {Borges Hink}, J.~M. {Beaver}, M.~A. {Buckner}, T.~{Morris},
  U.~{Adhikari}, and S.~{Pan}, ``Machine learning for power system disturbance
  and cyber-attack discrimination,'' in \emph{ISRCS}, 2014, pp. 1--8.

\bibitem{7805317}
U.~{Adhikari}, T.~H. {Morris}, and S.~{Pan}, ``Applying hoeffding adaptive
  trees for real-time cyber-power event and intrusion classification,''
  \emph{IEEE Transactions on Smart Grid}, vol.~9, no.~5, pp. 4049--4060, 2018.

\bibitem{musleh2019survey}
A.~S. Musleh, G.~Chen, and Z.~Y. Dong, ``A survey on the detection algorithms
  for false data injection attacks in smart grids,'' \emph{IEEE Transactions on
  Smart Grid}, vol.~11, no.~3, pp. 2218--2234, 2019.

\bibitem{cui2020detecting}
L.~Cui, Y.~Qu, L.~Gao, G.~Xie, and S.~Yu, ``Detecting false data attacks using
  machine learning techniques in smart grid: A survey,'' \emph{Journal of
  Network and Computer Applications}, p. 102808, 2020.

\bibitem{case2016analysis}
{Defence Use Case}, ``Analysis of the cyber attack on the {U}krainian power
  grid,'' 2016.

\bibitem{farwell2011stuxnet}
J.~P. Farwell and R.~Rohozinski, ``Stuxnet and the future of cyber war,''
  \emph{Survival}, vol.~53, no.~1, pp. 23--40, 2011.

\bibitem{hu2019advanced}
C.~Hu, J.~Yan, and C.~Wang, ``Advanced cyber-physical attack classification
  with extreme gradient boosting for smart transmission grids,'' in \emph{IEEE
  Power \& Energy Society General Meeting}.\hskip 1em plus 0.5em minus
  0.4em\relax IEEE, 2019, pp. 1--5.

\bibitem{cheng2020powernet}
Y.~Cheng, C.~Xu, D.~Mashima, P.~P. Biswas, G.~Chipurupalli, B.~Zhou, and Y.~Wu,
  ``Powernet: a smart energy forecasting architecture based on neural
  networks,'' \emph{IET Smart Cities}, vol.~2, no.~4, pp. 199--207, 2020.

\bibitem{biswas2019synthesized}
P.~P. Biswas, H.~C. Tan, Q.~Zhu, Y.~Li, D.~Mashima, and B.~Chen, ``A
  synthesized dataset for cybersecurity study of iec 61850 based substation,''
  in \emph{IEEE SmartGridComm}.\hskip 1em plus 0.5em minus 0.4em\relax IEEE,
  2019, pp. 1--7.

\bibitem{5128907}
H.~He and E.~A. Garcia, ``Learning from imbalanced data,'' \emph{IEEE TKDE},
  vol.~21, no.~9, pp. 1263--1284, 2009.

\bibitem{5504793}
R.~Sommer and V.~Paxson, ``Outside the closed world: On using machine learning
  for network intrusion detection,'' in \emph{IEEE Symposium on Security and
  Privacy}, 2010, pp. 305--316.

\bibitem{lixia2010accuracy}
M.~Lixia, C.~Muscas, and S.~Sulis, ``On the accuracy specifications of phasor
  measurement units,'' in \emph{I2MTC}.\hskip 1em plus 0.5em minus 0.4em\relax
  IEEE, 2010, pp. 1435--1440.

\bibitem{en12122272}
H.~León, C.~Montez, O.~Valle, and F.~Vasques, ``Real-time analysis of
  time-critical messages in {IEC} 61850 electrical substation communication
  systems,'' \emph{Energies}, vol.~12, no.~12, 2019.

\bibitem{Hoyos2012ExploitingTG}
J.~Hoyos, M.~Dehus, and T.~X. Brown, ``Exploiting the {GOOSE} protocol: A
  practical attack on cyber-infrastructure,'' \emph{2012 IEEE Globecom
  Workshops}, pp. 1508--1513, 2012.

\end{thebibliography}
\end{small}

\end{document}